\documentclass[12pt,preprint]{aastex}

\usepackage{epsfig,amssymb}
\usepackage{graphicx}
\usepackage{rotating}
\usepackage{mathrsfs}
\newcommand{\ben}{\begin{enumerate}}
\newcommand{\een}{\end{enumerate}}
\newcommand{\bfig}{\begin{figure}}
\newcommand{\efig}{\end{figure}}
\newcommand{\beq}{\begin{equation}}
\newcommand{\eeq}{\end{equation}}

\newcommand{\mcal}{\mathcal}

\def\degr{\hbox{$^{\circ}$}}

\shorttitle{Energy-Helicity Diagram of Solar Active Regions}

\shortauthors{Tziotziou, Georgoulis, \& Raouafi}

\begin{document}

\title{The Magnetic Energy -- Helicity Diagram of Solar Active Regions}

\author{Kostas Tziotziou \& Manolis K. Georgoulis\altaffilmark{1}}
\affil{Research Center for Astronomy and Applied Mathematics (RCAAM)\\
       Academy of Athens, 4 Soranou Efesiou Street, Athens, Greece, GR-11527}
\and
\author{Nour-Eddine Raouafi}
\affil{The Johns Hopkins University Applied Physics Laboratory (JHU/APL)\\
       11100 Johns Hopkins Rd. Laurel, MD 20723-6099, USA}

\altaffiltext{1}{Marie Curie Fellow.}

\begin{abstract}
Using a recently proposed nonlinear force-free method designed for single
vector magnetograms of solar active regions we calculate the instantaneous
free magnetic energy and relative magnetic helicity budgets in 162 vector
magnetograms corresponding to 42 different active regions. We find a
statistically robust, monotonic correlation between the free magnetic energy
and the relative magnetic helicity in the studied regions. This correlation
implies that magnetic helicity, besides free magnetic energy, may be an
essential ingredient for major solar eruptions. Eruptive active regions
appear well segregated from non-eruptive ones in {\em both} free energy and
relative helicity with major (at least M-class) flares occurring in active
regions with free energy and relative helicity exceeding $4 \times 10^{31}$
erg and $2 \times 10^{42}$ Mx$^2$, respectively. The helicity threshold
agrees well with estimates of helicity contents of typical coronal mass
ejections.
\end{abstract}


\section{Introduction}
\label{intro}

Active regions (ARs) are formed when considerable, localized magnetic flux
emergence occurs into the solar atmosphere, with characteristic fluxes of
the order of 10$^{22}$ Mx \citep{schr:harv}. ARs are magnetic structures far
from a ground, current-free (potential) energy state and, as such, they
store large amounts of free magnetic energy. Over the last two decades,
multiple reports highlight the simultaneous accumulation of large amounts of
magnetic helicity in ARs \citep[e.g.,][]{lab07, smyr10}. Magnetic helicity
emerges via helical magnetic flux tubes or is being generated by solar
differential rotation and peculiar photospheric motions and is a way to
quantify the stress and distortion of the magnetic field compared to its
potential-energy state. The free magnetic energy release is fragmented in
ARs and fuels solar flares and/or coronal mass ejections (CMEs) that tend to
relax the magnetic configuration. Contrary to magnetic energy, helicity
cannot be efficiently removed by magnetic reconnection \citep{berg84}. If it
is not transferred to larger scales in the Sun via existing magnetic
connections, it can only be expelled in the form of CMEs \citep{low94,
devo:00}. As a result, an isolated, confined magnetic configuration with
accumulated magnetic helicity cannot relax to a potential-field
configuration.

Although the importance of storage and release of free magnetic energy in
ARs is widely acknowledged for solar eruptions \citep[e.g.,][]{schr09}, the
role of magnetic helicity is still under debate. Indeed, it has been
demonstrated that helicity is not necessary for eruptions to occur
\citep{phil05,zucc09}. On the other hand, observational and modeling works
have shown that ARs with large {\it dominant} left- or right-handed helicity
give rise to more and/or major eruptions
\citep[e.g.,][]{nind04,tor05,lab07,nind09,geo09}. Instrumental to this
debate is the lack of robust methods to calculate the (relative to a
reference potential field) instantaneous magnetic helicity {\it budget} in
ARs. Existing methods basically restrict to either integrating in time the
relative helicity injection rate \citep{berfie84} or evaluating the relative
helicity formula \citep{finn85, berg99} in a volume by means of a
three-dimensional magnetic field extrapolation and a respective
gauge-dependent expression for the vector potential. The helicity injection
rate depends on the photospheric velocity field whose inference involves
significant uncertainties \citep[e.g.,][]{wels07} while the ``volume''
helicity calculation depends on the model-dependent nonlinear force-free
field extrapolation, also subject to uncertainties and ambiguities
\citep[e.g.,][ and references therein]{schr06,metc08}.

Recently, \citet{geo12} (hereafter GTR12) proposed a general force-free
approach to self-consistently calculate the {\em instantaneous} magnetic
free energy and relative helicity budgets from (photospheric or
chromospheric) vector magnetograms of solar ARs. The method does not rely on
any magnetic field extrapolation but it uses a magnetic connectivity matrix
that may be inferred by extrapolations, among other methods. As in GTR12, we
use in this work a simulated annealing method that converges to a {\it
unique} connectivity matrix to calculate the free magnetic energy and
relative magnetic helicity for a large sample of active-region vector
magnetograms, seeking a statistically robust correlation between these
physical parameters, if any. Section 2 briefly describes the method, its
application and results are discussed in Section 3, while Section 4
summarizes our findings.

\section{Nonlinear force-free magnetic energy and helicity budgets in solar ARs}

In GTR12 we combined the linear force-free technique of \citet{geo07} with
the properties of the mutual helicity as discussed by \citet{dem06} to
derive the instantaneous nonlinear force-free (NLFF) field energy and
helicity budgets in an AR. In particular, a vector magnetogram of a given AR
is translated into a collection of slender force-free flux tubes with known
footpoints, flux contents, and variable force-free parameters. These flux
tubes are defined after determining the magnetic connectivity matrix in the
AR's magnetogram, which provides the flux contents committed to the
connection between opposite-polarity flux partitions. The method of choice
to calculate this matrix was introduced by \citet{geo:rus} and was revised
by GTR12. It uses a simulated annealing method that globally (within the
field of view) minimizes the connection lengths, at the same time
guaranteeing that only opposite-polarity flux elements will be connected to
each other. This approach emphasizes complex active regions with intense
magnetic polarity inversion lines (PILs) in their lower boundaries. As
explained in GTR12, the method gives unique results, contrary to
model-dependent NLFF field extrapolations.

For a collection of $N$ slender flux tubes the free magnetic energy $E_c$ is
the sum of a self term $E_{c_{self}}$, due to the internal twist and writhe
of each flux tube, and a mutual term $E_{c_{mut}}$, due to interactions
between different flux tubes. GTR12 provide a {\it lower limit} of the free
energy $E_c$ for a given connectivity that (i) assumes no winding of a given
flux tube around others, and (ii) neglects contributions from potential flux
tubes induced by the existing tubes in a space-filling, force-free magnetic
configuration. This expression for $E_c$ reads
\begin{equation}
E_c = E_{c_{self}} + E_{c_{mut}} = A d^2 \sum _{l=1}^N \alpha _l^2 \Phi_l^{2
\delta} +
      {{1} \over {8 \pi}} \sum _{l=1}^N \sum _{m=1, l \ne m}^N
           \alpha _l \mcal{L}_{lm}^{arch} \Phi_l \Phi_m\;\;.
\label{Ec_fin}
\end{equation}
In Equation~(\ref{Ec_fin}) $A$ and $\delta$ are known fitting constants, $d$
is the pixel size of the magnetogram and $\Phi_l$ and $\alpha_l$ are the
respective flux and force-free parameter of flux tube $l$.
$\mcal{L}_{lm}^{arch}$ is the mutual-helicity factor of two arch-like (not
winding around each other) flux tubes. Inference of this factor was first
discussed by \citet{dem06} and was later refined by GTR12. Refinement
included the case of a ``matching" photospheric footpoint in a pair of flux
tubes. ``Matching" footpoint means that the like-polarity footpoints of a
given flux-tube pair are within the same magnetic partition; therefore, they
are unresolved (considered as coinciding) by the method. To calculate
$\mcal{L}_{lm}^{arch}$ one needs the relative footpoint locations of flux
tubes $l$ and $m$ and an assessment of whether $l$ is ``above" $m$, or vice
versa. For each of these cases one calculates a different
$\mcal{L}_{lm}^{arch}$-value in cases of ``matching" footpoint and
intersecting footpoint segments for the flux-tube pair. By definition, these
two values have opposite signs. In case of non-intersecting segments, the
two possible $\mcal{L}_{lm}^{arch}$-values collapse to a single value.
$\mcal{L}_{lm}^{arch}$ is always a real number with an absolute value lower
than one. The selected $\mcal{L}_{lm}^{arch}$ value is the one that assigns
a positive increment to the overall free magnetic energy because of the
flux-tube pair. For a pair ($l$,$m$) of flux tubes with force-free
parameters $\alpha_l$ and $\alpha_m$, respectively, this means $(\alpha_l +
\alpha_m) \mcal{L}_{lm}^{arch} > 0$. $\mcal{L}_{lm}^{arch}$ is assumed equal
to zero if the respective energy increment can only be negative (this
applies exclusively to the non-intersecting-segments case), which is not a
physical solution.

%Moreover, $\mcal{L}_{lm}^{close}$ is the Gauss linkage number determining
%the intertwining between interacting flux tubes. To keep $E_c$ minimum,
%$\mcal{L}_{lm}^{close}=0$, in case the lower-boundary footpoint segments of
%flux tubes $l$ and $m$ do not interact or in case $|\alpha_{lm}|$ is smaller
%than its uncertainty $\delta\alpha_{lm} = {{1} \over {2}}
%\sqrt{\delta\alpha_l^2 + \delta\alpha_m^2}$, and $|\mcal{L}_{lm}^{close}|=1$
%otherwise. The sign of $\mcal{L}_{lm}^{close}$ is determined by the sign of
%$\alpha_{lm}$ such that $\alpha _{lm} \mcal{L}_{lm}^{close}>0$, thus always
%contributing a positive increment in the free energy $E_c$ in case of
%interacting flux tubes.

The respective relative magnetic helicity $H_m$ for the collection of $N$
slender flux tubes is also the sum of a self ($H_{m_{self}}$) and a mutual
($H_{m_{mut}}$) term. GTR12 derive
\begin{equation}
H_m = H_{m_{self}} + H_{m_{mut}} = 8 \pi d^2 A \sum _{l=1}^N \alpha _l
\Phi_l ^{2 \delta} +
      \sum _{l=1}^N \sum _{m=1,l \ne m}^N \mcal{L}_{lm}^{arch} \Phi_l
      \Phi_m\;\;.
\label{Hm_fin}
\end{equation}
A detailed derivation of uncertainties for both $E_c$ and $H_m$ is also
provided in GTR12.

\section{Application to observed solar active region magnetic fields}
\label{obs}

\subsection{Data selection}

We have collected an extended sample of 162 photospheric and low
chromospheric vector magnetograms obtained by the ground-based Imaging
Vector Magnetograph \citep[IVM;][]{mic96,lab99} of the University of
Hawaii's Mees Solar Observatory and by the space-based Spectropolarimeter
\citep[SP; see description in][]{lite08} of the Solar Optical Telescope
(SOT) onboard {\em Hinode}. The IVM provides complete Stokes profiles of the
\ion{Fe}{1} 630.25 nm photospheric line (for earlier data) and of the
\ion{Na}{1} D 1 589.60 nm chromospheric line \citep{leka03}, for data
recorded after 2003, with a spatial sampling of 0.55 arcsec per pixel (full)
or 1.1 arcsec per pixel (binned). The SP provides full Stokes profiles of
the \ion{Fe}{1} 630.25/630.15 nm lines with a maximum spatial sampling of
0.16 arcsec per pixel. IVM magnetogram edges were carefully cropped to
remove instrumental border artifacts. To remove the intrinsic 180\degr\
azimuthal ambiguity in the vector magnetograms we applied the non-potential
field calculation (NPFC) method of \citet{geo05}, as revised in
\citet{metc06}. As typical uncertainties for the line-of-sight and
transverse field components we used ($\delta B_l$, $\delta B_{tr}$ )=(50,
100) G, for IVM data, and ($\delta B_l$, $\delta B_{tr}$ )=(5, 50) G, for
SOT/SP data.

Our sample of 162 vector magnetograms corresponds to 42 different ARs,
distinguished between ``non-flaring'' and ``flaring'' ones. The first
category involves 18 ARs that have not hosted a higher than C-class flare,
while the second involves 24 ARs with at least one M-class flare.
Non-flaring ARs include NOAA ARs 8844, 9114, 9635, 9845, 10050, 10254,
10323, 10349, 10536, 10939, 10940, 10953, 10955, 10956, 10961, 10963, 10971,
and 10978. Flaring ARs include NOAA ARs 8210, 9026, 9165, 9393, 9415, 9632,
9661, 9684, 9704, 9773, 10030, 10039, 10162, 10365, 10375, 10386, 10484,
10488, 10501, 10570, 10596, 10656, 10930, and 10960. For a few ARs we have
timeseries of vector magnetograms spanning from a few hours of IVM data
(e.g., NOAA ARs 8844, 9165) to a few days of SOT/SP data (e.g., NOAA ARs
10930, 10956). Moreover, our sample includes both short-lived emerging flux
regions (i.e., NOAA AR 8844) and persistent, large and complex regions
(e.g., NOAA ARs 10488, 10930).

\begin{figure}
\includegraphics[width=\linewidth]{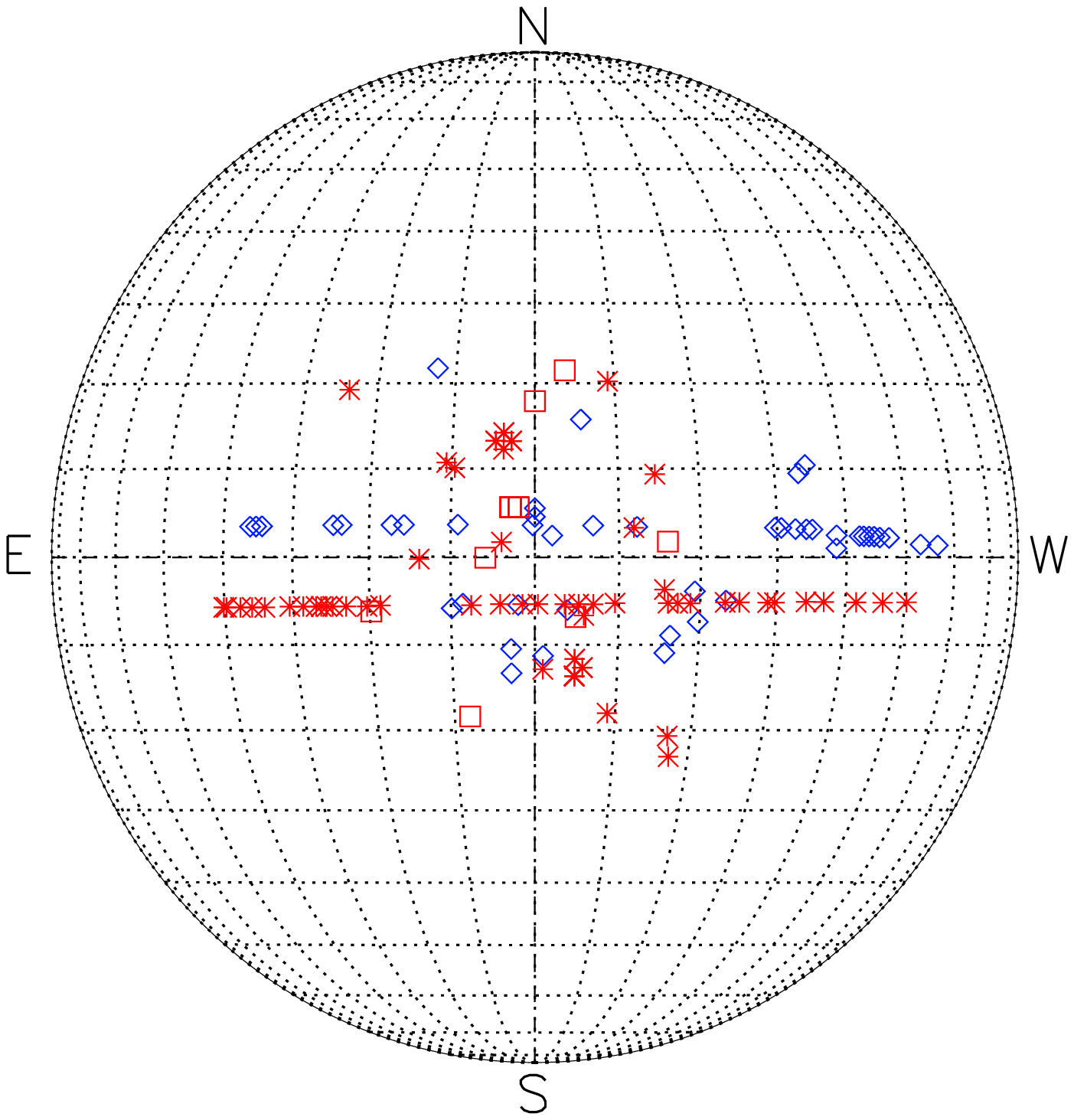}
\caption{Central heliographic positions for our 162 vector magnetograms.
Blue diamonds correspond to non-flaring (up to C-class flaring) ARs while
red squares and asterisks correspond to M- and X-class flaring ARs,
respectively.} \label{noaa}
\end{figure}

Figure~\ref{noaa} shows the central heliographic position of the selected
162 magnetograms. Locations cover a latitudinal area of $\pm$25\degr\ and a
meridional zone of $-40$\degr\ to $+60$\degr\ . Larger central-meridian
distances were avoided to avoid extreme projection effects. The calculations
below involve the local, heliographic field components on the image plane of
each magnetogram.
%As for the sign of relative helicity, there seems to be no
%hemispheric dependence as the percentage of Northern hemisphere magnetograms
%showing positive (negative) helicity is 50\% (50\%), while the respective
%Southern hemisphere percentage is 42\% (58\%).

\subsection{Magnetic energy -- relative magnetic helicity budgets and their relation}

\begin{figure*}
\includegraphics[width=\linewidth]{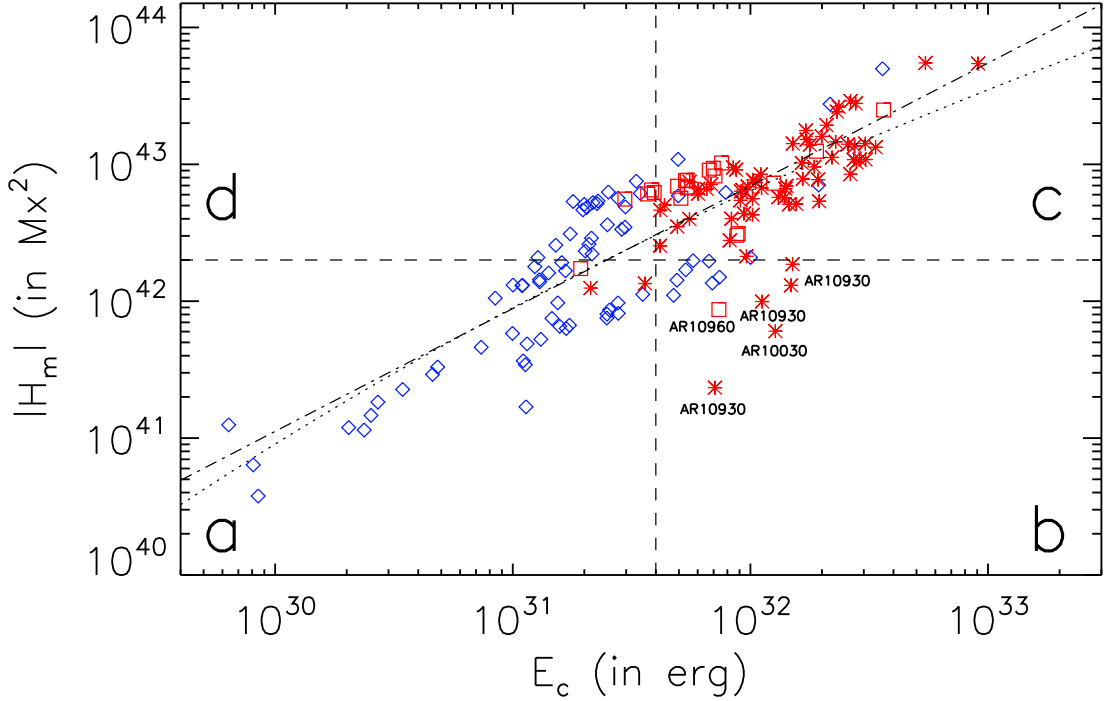}
\caption{The free energy -- relative helicity diagram of solar ARs. Blue
diamonds, red squares and red asterisks correspond to non-flaring, M- and
X-class flaring ARs, respectively. Dashed lines indicate the estimated
thresholds for relative magnetic helicity ($\sim 2 \times 10^{42}$ Mx$^2$)
and free magnetic energy ($\sim 4 \times 10^{31}$ erg) above which ARs seem
to give major flares almost exclusively. These thresholds divide the diagram
in four Regions, labeled {\em a}, {\em b}, {\em c} and {\em d} (see text).
The dotted and dash-dotted lines denote the least-squares best fit (Equation
(\ref{scale1})) and the least-squares best logarithmic fit (Equation
(\ref{scale2})), respectively. Some NOAA AR numbers are also indicated in
Figure~\ref{enhelerr} and are discussed in the text.} \label{enhel}
\end{figure*}

Using equations (\ref{Ec_fin}) and (\ref{Hm_fin}) we calculate the free
magnetic energy and relative magnetic helicity budgets for the selected 162
vector magnetograms. The results suffice to construct the {\it free-energy
-- relative helicity diagram} (hereafter energy - helicity [EH] diagram) of
solar ARs, shown in Figure~\ref{enhel}. Two main conclusions stem from this
plot: first, there is a nearly monotonic dependence of direct
proportionality between $E_c$ and $H_m$; flaring ARs tend to show {\it both}
large free energies and large amplitudes of relative helicity. Second,
flaring and non-flaring ARs appear well segregated; excesses of $\sim 4
\times 10^{31}$ erg in free magnetic energy {\it and} of $\sim 2 \times
10^{42}$ Mx$^2$ in relative magnetic helicity tend to bring ARs into major
flaring (or eruptive, in general), territory. Within this area, there is no
appreciable segregation between M- and X-flaring ARs. Later in this Section
we discuss the (few) exceptions to the above general assessment. A similar
diagram (not shown) between the free magnetic energy and the unsigned
(total) magnetic flux shows significantly less segregation between flaring
and non-flaring ARs than that of the EH diagram.

The least-squares best fit between $|H_m|$ and $E_c$ reveals a scaling of
the form
\begin{equation}
\log |H_{\rm m}| \propto 53.4 - 0.0524 \, (\log E_{\rm c})^{0.653} \, \exp
\frac{97.45}{\log E_{\rm c}}\;\;, \label{scale1}
\end{equation}
while, in general, a simpler logarithmic scaling of the form
\begin{equation}
|H_{\rm m}| \propto 1.37 \times 10^{14} E_c^{0.897} \label{scale2}
\end{equation}
also works quite well. Both have a goodness of fit $\sim 0.7$ with the
Kolmogorov-Smirnov test giving also similar significance levels.
Equation~(\ref{scale1}) implies that relative magnetic helicity builds up
with faster fractional rates than magnetic free energy in the initial stages
of active regions, however this is not conclusively supported by the derived
fit.

\begin{figure*}
\includegraphics[width=\linewidth]{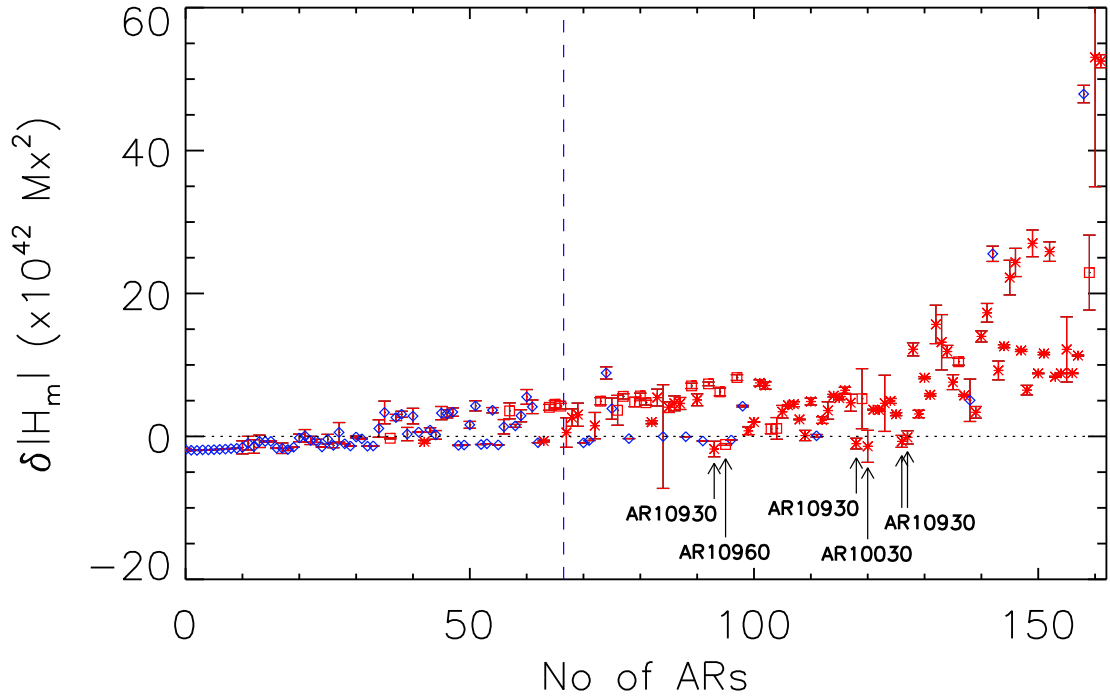}
\caption{Uncertainties for the difference between the relative helicity
magnitudes and the inferred relative helicity threshold of $2 \times
10^{42}$ Mx$^2$, ordered by increasing free magnetic energy. The dashed
vertical line separates ARs below (left) and above (right) the respective
free magnetic energy threshold. Symbols and marked NOAA AR numbers are the
same with those of Figure~\ref{enhel}. } \label{enhelerr}
%\caption{Uncertainties for the relative helicity magnitudes, ordered by
%increasing free magnetic energy. The dotted line denotes the relative
%helicity magnitude threshold of $2 \times 10^{42}$ Mx$^2$
%(Figure~\ref{enhel}) while the dashed vertical line separates ARs below
%(left) and above (right) the respective threshold for the free magnetic
%energy ($4 \times 10^{31}$ erg). Marked NOAA AR numbers are also indicated
%in Figure~\ref{enhel} and are discussed in the text. } \label{enhelerr}
\end{figure*}

Figure~\ref{enhelerr} shows the calculated uncertainties for the helicity
magnitude of the studied 162 magnetograms, ordered by increasing free
magnetic energy. The respective uncertainties for the free magnetic energy
are not shown because they are quite smaller: below the free energy
threshold the mean error is $\sim$22\% while above it the mean error becomes
$\sim$7\%; hence, they do not affect significantly the EH diagram of
Figure~\ref{enhel}, especially its high-energy part. Uncertainties in the
relative helicity are higher: below the relative helicity threshold the mean
error is $\sim$50\% while above it the mean error drops to $\sim$14\%.
%This explains the observed dispersion of points in Region {\em a}, mainly,
%and Region {\em b} of Figure~\ref{enhel}, to a smaller extent. As helicity
%magnitudes increase further, fractional uncertainties seem to decrease
%statistically.
As with free-energy uncertainties, the respective helicity uncertainties do
not alter significantly the EH diagram of Figure~\ref{enhel}. Point taken,
there are some notable exceptions (see below).

Before discussing exceptions, let us briefly discuss how the two thresholds
in free energy and relative helicity divide the EH diagram of
Figure~\ref{enhel}. Regions {\em a} and {\em c} include populations of small
$E_c$ / small $|H_m|$ and large $E_c$ / large $|H_m|$, respectively and hold
the vast majority of points in the diagram, thus reflecting the nearly
monotonic free-energy -- helicity dependence. Region {\em b} indicates ARs
with a large free energy but with a small, or relatively small, relative
helicity budget. Given that magnetic helicity is a signed quantity, a
population in Region {\em b} might also include ARs with significant, but
similar, amounts of both senses of helicity. Other than four (4)
magnetograms of X-flaring ARs and one (1) magnetogram of a M-flaring AR
(discussed below) there is no clear population in Region {\em b}. This is
important evidence to the existence of a {\it dominant} sense of helicity in
strongly helical ARs. Finally, Region {\em d} indicates ARs with small, or
relatively small, free energy and a large relative helicity budget. A
population in this Region would be troublesome to interpret or would
indicate a problematic numerical approach to calculate $E_c$ and $|H_m|$.
Fortunately, other than some typical scatter in the EH diagram, close to
both thresholds, there is no population in Region {\em d}.

We now discuss some notable exceptions to the physical interpretation of the
EH diagram of Figure~\ref{enhel}: first, four magnetograms of flaring NOAA
AR 9165 in Region {\em d} are within uncertainties from the energy
threshold, hence, they are not considered exceptions. In Region {\em c}
there are 7 magnetograms of non-flaring ARs (NOAA ARs 10963, 9845, 10953,
10978, 10536, 10323 and 10349, by increasing free magnetic energy) that
clearly go well into major-flaring territory. Most of these ARs share a
common feature: rather than hosting one or more major flares (they were
apparently capable for this based on their free magnetic energy and helicity
budgets) they released substantial free energy in lengthy series of C-class
flares. For example, NOAA AR 10536 produced 34 C-class flares within two
weeks (the highest being a C7.7 flare) while NOAA AR 10349 produced 36
C-class flares in ten days. Two magnetograms (NOAA ARs 10978 and 10323)
where recorded after a C-class flare.
%Two M-class flaring ARs (NOAA ARs 9704 and 10162) are also well into X-class
%flaring territory. They were both extremely active in C- and M-class flares
%with NOAA AR 10162 giving a practically X-class (M9.9) flare.
In Region {\em b} there are 4 magnetograms of X-flaring ARs (NOAA ARs 10030
and 10930) and one magnetogram of a M-flaring AR (NOAA AR 10960) that show
deficits in their relative helicity budgets. All X-flaring ARs (marked in
Figure~\ref{enhelerr}) show relatively high uncertainties in relative
helicity, with fractional uncertainties in the range of 0.5 to 4.7, capable
of placing them in Region {\em c}. Moreover, and related to the large
uncertainties, we cannot rule out possible white-light contamination due to
flaring in these regions: data for the magnetogram of NOAA AR 10030 were
recorded right after two X and M-class flares while magnetograms of NOAA AR
10930 were recorded between swarms of B- and C-class flares. For the
M-flaring NOAA AR 10960 we cannot exclude the possibility of similar
helicity budgets of both senses. We cannot assume the same for previously
discussed NOAA ARs 10030 and 10930 because they participate in the sample
with timeseries of magnetograms that, excluding the ones in Region {\em b},
reside exclusively in Region {\em c}. Finally there is one magnetogram of a
M-flaring (NOAA AR 10501) and two of X-flaring (NOAA ARs 10375 and 10386)
ARs that fall short of the energy and helicity thresholds, residing into
Region {\it a}. Significant errors (both in free magnetic energy and
relative helicity) or contamination by white-light flare emission may also
be responsible for this effect (the NOAA AR 10375 magnetogram was recorded
during the declining phase of a C2.5 flare) but, moreover, it is known that
intense, localized magnetic flux emergence can also force major eruptive
flares \citep[e.g.,][]{nitta01, zhang02} before an AR manages to build a
strong PIL and, consequently, large budgets of $E_c$ and $H_m$.

%To test the robustness of the EH diagram of Figure~\ref{enhel} we have
%increased the threshold above which the Gauss linkage number
%$\mcal{L}_{lm}^{close}$ obtains a nonzero value in Equations (\ref{Ec_fin}),
%(\ref{Hm_fin}). GTR12 and other, previous works report that the mutual terms
%of free energy and helicity far exceed the respective self terms in
%amplitude. So far we have used a $1 \sigma$-threshold, i.e.
%$|\mcal{L}_{lm}^{close}=1|$ for interacting footpoint segments if
%$|\alpha_{lm}| > \delta\alpha_{lm}$. Imposing a $2 \sigma$-threshold, i.e.
%$|\mcal{L}_{lm}^{close}=1|$ if $|\alpha_{lm}| > 2 \delta\alpha_{lm}$, does
%not change the EH diagram significantly. A comparison between the $1 \sigma$
%and the $3 \sigma$ values for $|H_m|$ and $E_c$, provided in
%Figure~\ref{scatter}, shows that both $|H_m|$ and $E_c$ tend to decrease, as
%expected, because amounts of mutual helicity and free energy are removed,
%without however destroying the segregation between flaring and non-flaring
%ARs. We choose a $1 \sigma$-threshold because more stringent criteria
%increase the scatter in the EH diagram. We feel this is an artifact caused
%by the removal of large, existing, free energy and relative helicity
%budgets.

\begin{figure*}
\includegraphics[width=1.\linewidth]{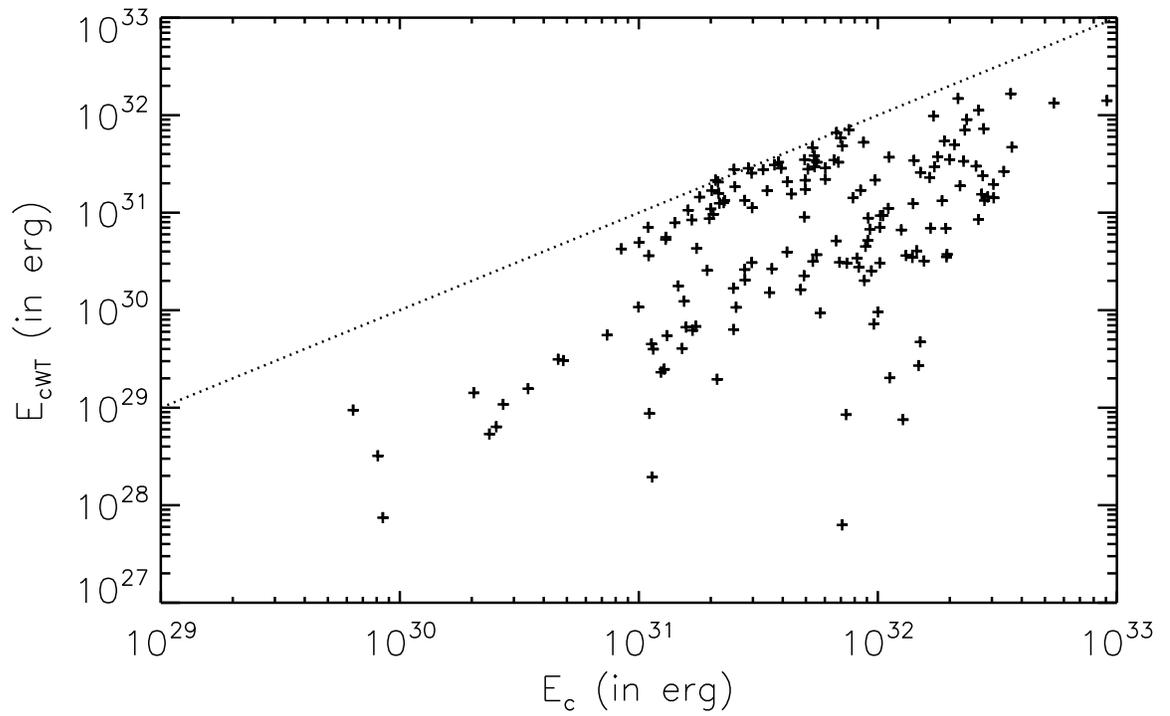}
\caption{Scatter plot for free magnetic energies $E_{c}$ and their
respective Woltjer-Taylor minima $E_{c_{WT}}$ (Equation~(\ref{Ec_wt})). The
dotted line denotes equality between the two energy estimates. }
\label{scatter}
\end{figure*}

To test the validity of our free magnetic energies $E_c$, we compare them
with the lowest possible free magnetic energy $E_{c_{WT}}$ that corresponds
to a given amount of relative helicity for the NLFF field. This limiting
value is the LFF free magnetic energy corresponding to this helicity, per
the Woltjer-Taylor theorem \citep{wol58, tay74, tay86}. As GTR12 have shown,
$E_{c_{WT}}$ is given by
\begin{equation}
E_{c_{WT}} = {{H_m^2} \over {(8 \pi d)^2 A \Phi ^{2 \lambda}}}\;\;.
\label{Ec_wt}
\end{equation}
All calculated free magnetic energies $E_c$ must be larger than this value.
As Figure~\ref{scatter} clearly demonstrates, this condition, within
applicable errors, is always satisfied.

\section{Conclusions}
\label{dis}

We have applied a new NLFF field method to calculate the instantaneous free
magnetic energy and relative magnetic helicity budgets of solar active
regions. On a sample of 162 such magnetograms we report, for the first time,
(1) a nearly monotonic relation of direct proportionality between the free
magnetic energy and the relative magnetic helicity in ARs and (2) the
existence of thresholds $\sim 4 \times 10^{31}$ erg and $\sim 2 \times
10^{42}$ Mx$^2$, for the free energy and the relative helicity,
respectively, for ARs to host major, typically eruptive, flares. As magnetic
helicity is a signed quantity, the monotonic dependence between it and the
free energy implies that, in spite of simulations reporting that helicity is
not necessary for solar eruptions (see Introduction), real flaring/eruptive
ARs show a significant budget and a dominant sense of magnetic helicity.
This finding
%, together with the
%nonlinear dependence between $|H_m|$ and $E_c$ with helicity increasing
%faster than free energy for small budgets (Region {\em a} in
%Figure~\ref{enhel})
appears to suggest an important role of magnetic helicity in solar
eruptions. However, the details of this role are yet to be uncovered.

An additional important finding refers to the inferred threshold values for
$E_c$ and $|H_m|$. While a threshold of $4 \times 10^{31}$ erg in free
energy suffices to justify at least one M-class flare \citep{huds11}, a
threshold of $2 \times 10^{42}$ Mx$^2$ in relative helicity is in excellent
agreement with estimated helicity budgets of typical CMEs ($\sim 2 \times
10^{42}$ Mx$^2$ by \citet{devo:00}; $(1.8 - 7) \times 10^{42}$ Mx$^2$ by
\citet{geo09}). This may indicate that active-region CMEs occur only when
source ARs can shed the required helicity budget for them, in line with
theoretical CME interpretations as means of relieving the Sun from its
excess helicity.

The above results are a first step to assess and uncover the possible {\it
combined} role of free magnetic energy and relative magnetic helicity in
solar eruptions. We do not envision or expect use of these results for
eruption forecasting purposes at this stage. However, these findings
instigate focused research on the detailed, quantitative role of magnetic
helicity in solar eruptions that could even lead to the formulation of
credible eruption initiation mechanisms. We intend to investigate such
possible physical links in the future.

\acknowledgments

We gratefully acknowledge the Institute of Astronomy and Astrophysics, Space
Applications and Remote Sensing of the National Observatory of Athens for
the availability of their computing cluster facility. IVM Survey data are
made possible through the staff of the U. of Hawaii. Hinode is a Japanese
mission developed and launched by ISAS/JAXA, with NAOJ as domestic partner
and NASA and STFC (UK) as international partners and operated in
co-operation with ESA and NSC (Norway). This work was partially supported
from NASA's Guest Investigator grant NNX08AJ10G and from the EU's Seventh
Framework Programme under grant agreement n$^o$ PIRG07-GA-2010-268245.

\clearpage

%\begin{figure*}
%\includegraphics[width=1.\linewidth]{fig5.ps}
%\caption{Flux contents flux .} \label{flux}
%\end{figure*}


\begin{thebibliography}{}

\bibitem[Berger(1984)]{berg84}
Berger, M.~A.\ 1984, Ph.D.~Thesis

\bibitem[Berger \& Field(1984)]{berfie84}
Berger, M.~A., \& Field, G.~B.\ 1984, Journal of Fluid Mechanics, 147, 133

\bibitem[Berger(1999)]{berg99}
Berger, M.~A.\ 1999, Plasma Physics and Controlled Fusion, 41, 167

\bibitem[Demoulin et al.(2006)]{dem06}
Demoulin, P., Pariat, E., \& Berger, M. A. 2006, \solphys, 233, 3

\bibitem[DeVore(2000)]{devo:00}
DeVore, C.~R. 2000, \apj, 539, 944

\bibitem[Finn \& Antonsen(1985)]{finn85}
Finn, J.~M., \& Antonsen, T.~M., Jr.\ 1988, Plasma Phys. Controlled Fusion,
9, 111

\bibitem[Georgoulis(2005)]{geo05}
Georgoulis, M. K. 2005, \apj, 629, L69

\bibitem[Georgoulis \& LaBonte(2007)]{geo07}
Georgoulis, M. K., \& LaBonte, B. J. 2007, \apj, 671, 1034

\bibitem[Georgoulis \& Rust(2007)]{geo:rus}
Georgoulis, M. K., \& Rust, D. M. 2007, \apj, 661, L109

\bibitem[Georgoulis et al.(2009)]{geo09}
Georgoulis, M.~K., Rust, D.~M., Pevtsov, A.~A., Bernasconi, P.~N., \&
Kuzanyan, K.~M. 2009, \apjl, 705, L48

\bibitem[Georgoulis et al.(2012)]{geo12}
Georgoulis, M. K., Tziotziou, K., \& Raouafi, N.-E. 2012, \apj, in press
(GTR12)

\bibitem[Hudson(2011)]{huds11}
Hudson, H.~S.\ 2011, \ssr, 158, 5

\bibitem[LaBonte et al.(1999)]{lab99}
LaBonte, B. J., Mickey, D. L., \& Leka, K. D. 1999, \solphys, 189, 1

\bibitem[LaBonte et al.(2007)]{lab07}
LaBonte, B.~J., Georgoulis, M.~K., \& Rust, D.~M.\ 2007, \apj, 671, 955

\bibitem[Leka \& Metcalf(2003)]{leka03}
Leka, K.~D., \& Metcalf, T.~R.\ 2003, \solphys, 212, 361

\bibitem[Lites et al.(2008)]{lite08}
Lites, B.~W., Kubo, M., Socas-Navarro, H., et al.\ 2008, \apj, 672, 1237

\bibitem[Low(1994)]{low94}
Low, B.~C.\ 1994, Physics of Plasmas, 1, 1684

\bibitem[Metcalf et al.(2006)]{metc06}
Metcalf, T.~R., Leka, K.~D., Barnes, G., et al.\ 2006, \solphys, 237, 267

\bibitem[Metcalf et al.(2008)]{metc08}
Metcalf, T.~R., De Rosa, M.~L., Schrijver, C.~J., et al.\ 2008, \solphys,
247, 269

\bibitem[Mickey et al.(1996)]{mic96}
Mickey, D. L., Canfield, R. C., LaBonte, B. J., Leka, K. D., Waterson, M.
F., \& Weber, H. M. 1996, \solphys, 168, 229

\bibitem[Nindos et al.(2003)]{nind:03}
Nindos, A., Zhang, J., \& Zhang, H. 2003, \apj, 594, 1033

\bibitem[Nindos \& Andrews(2004)]{nind04}
Nindos, A., \& Andrews, M.~D.\ 2004, \apjl, 616, L175

\bibitem[Nindos(2009)]{nind09}
Nindos, A.\ 2009, IAU Symposium, 257, 133

\bibitem[Nitta \& Hudson(2001)]{nitta01} Nitta, N.~V., \& Hudson, H.~S.\ 2001,
\grl, 28, 3801

\bibitem[Phillips et al.(2005)]{phil05}
Phillips, A.~D., MacNeice, P.~J., \& Antiochos, S.~K.\ 2005, \apjl, 624,
L129

\bibitem[Schrijver(2009)]{schr09}
Schrijver, C.~J.\ 2009, Advances in Space Research, 43, 739

\bibitem[Schrijver \& Harvey(1994)]{schr:harv}
Schrijver, C.~J., \& Harvey, K.~L. 1994, \solphys, 150, 1

\bibitem[Schrijver et al.(2006)]{schr06}
Schrijver, C.~J., De Rosa, M.~L., Metcalf, T.~R., et al.\ 2006, \solphys,
235, 161

\bibitem[Smyrli et al.(2010)]{smyr10}
Smyrli, A., Zuccarello, F., Romano, P., et al.\ 2010, \aap, 521, A56

\bibitem[Taylor(1974)]{tay74} Taylor, J.~B.\ 1974, Physical
Review Letters, 33, 1139

\bibitem[Taylor(1986)]{tay86} Taylor, J.~B.\ 1986, Reviews of
Modern Physics, 58, 741

\bibitem[T{\"o}r{\"o}k \& Kliem(2005)]{tor05}
T{\"o}r{\"o}k, T., \& Kliem, B.\ 2005, \apjl, 630, L97

\bibitem[Woltjer(1958)]{wol58} Woltjer, L.\ 1958, Proceedings
of the National Academy of Science, 44, 489

\bibitem[Welsch et al.(2007)]{wels07}
Welsch, B.~T., Abbett, W.~P., De Rosa, M.~L., et al.\ 2007, \apj, 670, 1434

\bibitem[Zhang \& Wang(2002)]{zhang02}
Zhang, J., \& Wang, J.\ 2002, \apjl, 566, L117

\bibitem[Zuccarello et al.(2009)]{zucc09}
Zuccarello, F.~P., Jacobs, C., Soenen, A., et al.\ 2009, \aap, 507, 441

\end{thebibliography}
\end{document}